\documentclass[twocolumn]{aastex62}

\newcommand\apjcls{1}
\newcommand\aastexcls{2}
\newcommand\othercls{3}

\newcommand\papercls{\aastexcls}
\if\papercls \apjcls
\usepackage{apjfonts}
\else\if\papercls \othercls
\usepackage{epsfig}
\usepackage{margin}
\usepackage{times}
\fi\fi
\usepackage{ifthen}
\usepackage{natbib}
\usepackage{bm}
\usepackage{amssymb, amsmath}
\usepackage{appendix}
\usepackage{etoolbox}
\usepackage[T1]{fontenc}
\usepackage{paralist}
\usepackage{newtxtext,newtxmath}
\if\papercls \apjcls
\newcommand\aas{\ref@jnl{AAS Meeting Abstracts}}
\newcommand\dps{\ref@jnl{AAS/DPS Meeting Abstracts}}
\newcommand\maps{\ref@jnl{MAPS}}
\else\if\papercls \othercls
\usepackage{astjnlabbrev-jh}
\fi\fi

\if\papercls \aastexcls
\hypersetup{citecolor=blue, 
            linkcolor=blue, 
            menucolor=blue, 
            urlcolor=blue}  
\else
\usepackage[
bookmarks=true,           
bookmarksnumbered=true,   
colorlinks=true,          
citecolor=blue,           
linkcolor=blue,           
menucolor=blue,           
urlcolor=blue,            
linkbordercolor={0 0 1},  
pdfborder={0 0 1},
frenchlinks=true]{hyperref}
\fi
\if\papercls \othercls

\else

\fi

\providecommand{\adsurl}[1]{\href{#1}{ADS}}

\makeatletter
\patchcmd{\NAT@citex}
  {\@citea\NAT@hyper@{%
     \NAT@nmfmt{\NAT@nm}%
     \hyper@natlinkbreak{\NAT@aysep\NAT@spacechar}{\@citeb\@extra@b@citeb}%
     \NAT@date}}
  {\@citea\NAT@nmfmt{\NAT@nm}%
   \NAT@aysep\NAT@spacechar\NAT@hyper@{\NAT@date}}{}{}

\patchcmd{\NAT@citex}
  {\@citea\NAT@hyper@{%
     \NAT@nmfmt{\NAT@nm}%
     \hyper@natlinkbreak{\NAT@spacechar\NAT@@open\if*#1*\else#1\NAT@spacechar\fi}%
       {\@citeb\@extra@b@citeb}%
     \NAT@date}}
  {\@citea\NAT@nmfmt{\NAT@nm}%
   \NAT@spacechar\NAT@@open\if*#1*\else#1\NAT@spacechar\fi\NAT@hyper@{\NAT@date}}
  {}{}
\makeatother

\makeatletter
\DeclareRobustCommand{\lowcase}[1]{\@lowcase#1\@nil}
\def\@lowcase#1\@nil{\if\relax#1\relax\else\MakeLowercase{#1}\fi}
\makeatother

\DeclareSymbolFont{UPM}{U}{eur}{m}{n}
\DeclareMathSymbol{\umu}{0}{UPM}{"16}
\let\oldumu=\umu
\renewcommand\umu{\ifmmode\oldumu\else\math{\oldumu}\fi}

\if\papercls \othercls

\else

\fi

\let\oldsim=\sim
\renewcommand\sim{\ifmmode\oldsim\else\math{\oldsim}\fi}
\let\oldpm=\pm
\renewcommand\pm{\ifmmode\oldpm\else\math{\oldpm}\fi}
\newcommand\by{\ifmmode\times\else\math{\times}\fi}

\newbox{\wdbox}
\renewcommand\c{\setbox\wdbox=\hbox{,}\hspace{\wd\wdbox}}
\renewcommand\i{\setbox\wdbox=\hbox{i}\hspace{\wd\wdbox}}

\newcount\timect
\newcount\hourct
\newcount\minct
\newcommand\now{\timect=\time \divide\timect by 60
         \hourct=\timect Cltiply\hourct by 60
         \minct=\time \advance\minct by -\hourct
         \number\timect:\ifnum \minct < 10 0\fi\number\minct}

\catcode`@=11

\newcommand\comment[1]{}

\newcommand\commenton{\catcode`\%=14}

\renewcommand\math[1]{$#1$}
\newcommand\mathshifton{\catcode`\$=3}

\let\atab=&
\newcommand\atabon{\catcode`\&=4}

\let\oldmsp=\sp
\let\oldmsb=\sb
\def\sp#1{\ifmmode
           \oldmsp{#1}%
         \else\strut\raise.85ex\hbox{\scriptsize #1}\fi}
\def\sb#1{\ifmmode
           \oldmsb{#1}%
         \else\strut\raise-.54ex\hbox{\scriptsize #1}\fi}
\newbox\@sp
\newbox\@sb
\def\sbp#1#2{\ifmmode%
           \oldmsb{#1}\oldmsp{#2}%
         \else
           \setbox\@sb=\hbox{\sb{#1}}%
           \setbox\@sp=\hbox{\sp{#2}}%
           \rlap{\copy\@sb}\copy\@sp
           \ifdim \wd\@sb >\wd\@sp
             \hskip -\wd\@sp \hskip \wd\@sb
           \fi
        \fi}
\def\msp#1{\ifmmode
           \oldmsp{#1}
         \else \math{\oldmsp{#1}}\fi}
\def\msb#1{\ifmmode
           \oldmsb{#1}
         \else \math{\oldmsb{#1}}\fi}

\def\supon{\catcode`\^=7}

\def\subon{\catcode`\_=8}

\def\supsubon{\supon \subon}

\newcommand\actcharon{\catcode`\~=13}

\newcommand\paramon{\catcode`\#=6}

\comment{And now to turn us totally on and off...}

\newcommand\reservedcharson{ \commenton  \mathshifton  \atabon  \supsubon 
                             \actcharon  \paramon}

\catcode`@=12
\reservedcharson

\if\papercls \apjcls

\else

\fi

\newcommand\chisq{\ifmmode{\chi\sp{2}}\else\math{\chi\sp{2}}\fi}
\newcommand\redchisq{\ifmmode{ \chi\sp{2}\sb{\rm red}}
                    \else\math{\chi\sp{2}\sb{\rm red}}\fi}
\newcommand\Teq{\ifmmode{T\sb{\rm eq}}\else$T$\sb{eq}\fi}
\newcommand\mjup{\ifmmode{M\sb{\rm Jup}}\else$M$\sb{Jup}\fi}
\newcommand\rjup{\ifmmode{R\sb{\rm Jup}}\else$R$\sb{Jup}\fi}
\newcommand\msun{\ifmmode{M\sb{\odot}}\else$M\sb{\odot}$\fi}
\newcommand\rsun{\ifmmode{R\sb{\odot}}\else$R\sb{\odot}$\fi}
\newcommand\mearth{\ifmmode{M\sb{\oplus}}\else$M\sb{\oplus}$\fi}
\newcommand\rearth{\ifmmode{R\sb{\oplus}}\else$R\sb{\oplus}$\fi}

\begin{document}

\title{Self-Consistent Evolution Models Show Weak Double-Diffusive Mixing \\in Jupiter and Saturn}

\author{J. R. Fuentes}
\affiliation{\rm TAPIR, California Institute of Technology, Pasadena, CA 91125, USA}

\author{Ankan Sur}
\affiliation{\rm Department of Earth, Planetary, and Space Sciences, University of California, Los Angeles, CA 90095, USA}

\author{David J. Stevenson}
\affiliation{\rm Division of Geological and Planetary Sciences, California Institute of Technology, Pasadena, CA 91125, USA}

\author{Peter Bodenheimer}
\affiliation{\rm UCO/Lick Observatory, Department of Astronomy and Astrophysics, University of California, Santa Cruz, CA 95064, USA}

\begin{abstract}
Double-diffusive convection in the ``fuzzy'' cores of giant planets has been widely discussed as a mechanism for redistributing heavy elements, but its efficiency in evolutionary models remains uncertain. Here we implement a prescription for transport across a double-diffusive staircase in the planetary evolution code \texttt{APPLE} and apply it to post-formation interior models of Jupiter and Saturn that contain steep compositional gradients.  We find that after evolving the models for 4.56 Gyr, the transport associated with double-diffusive convection alone cannot transform steep composition gradients and compact cores into extended, partially mixed fuzzy cores. Our results show that in both Jupiter and Saturn, double-diffusive convection redistributes $\approx 0.1\,M_\oplus$ of heavy elements over the age of the Solar System in our fiducial models, and at most $\approx 2\,M_\oplus$ even when the effective diffusivities are enhanced by three orders of magnitude. We show that this inefficiency arises because the buoyancy work available to drive compositional transport is constrained by the small thermal energy budget of the deep interior, in contrast to idealized Boussinesq simulations that operate in regimes more favorable to layer merging and efficient mixing. Since the present-day interiors of Jupiter and Saturn exhibit substantially more extended heavy-element distributions, those structures must have established during formation through additional transport processes not captured by the formation models considered here.
\end{abstract}


\section{Introduction}

Understanding the internal structure of giant planets requires a consistent treatment of both their formation history and subsequent evolution. Current observational constraints on Jupiter and Saturn, coming from gravity data from the Juno and Cassini missions, together with ring-seismology measurements, are best matched by interior models with extended, partially mixed ``fuzzy'' cores rather than sharp, compact core-envelope structures \citep[e.g.,][]{Wahl2017,Mankovich2021}.

So far, two classes of models have been used to characterize the interior evolution of Jupiter and Saturn in light of the fuzzy core requirement. The first class of models evolves a primordial planet with initial conditions selected based on their ability to reproduce present-day observational constraints. Such initial conditions require a primordial planet with composition gradients that are already fuzzy or spatially extended, as well as relatively low-entropy (``cold'') interiors \citep[e.g.,][]{Tejada2025,Sur2025b,Sur2025, Knierim2026}. Because the available data alone do not uniquely constrain the deep interior structure of Jupiter and Saturn, this approach admits a broad range of viable present-day configurations \citep[e.g.,][]{Militzer2024}. In contrast, the second class of models is motivated by our current understanding of the physics of planet formation. These models predict substantially hotter post-formation interiors with steep compositional gradients, and after evolving them for 4.56 Gyr, often fail to reproduce the extent of the fuzzy core required by the observational constraints \citep[e.g.,][]{Muller2020, Stevenson2022, Bodenheimer2025}. Although the second class of models are more firmly grounded in the physical processes of accretion, their model assumptions, such as the efficiency of planetesimal dissolution and the distribution of solids during envelope growth, are themselves subject to substantial uncertainty.

The difference between the final outcome of these two classes of models motivates the question of whether the composition gradients produced at the end of planet formation can be substantially modified during the subsequent cooling and contraction of the planet. In particular, if an efficient transport mechanism operates after accretion, could it transform a steep heavy-element distribution into a more spatially extended configuration? We address this question by following the thermal and compositional evolution of giant planets, starting from currently available formation models in the literature.

The subsequent evolution of steep compositional gradients after the end of accretion is governed by the planet's loss of internal heat. The large energy outflow (IR radiation) as the planet cools and contracts drives vigorous convection in a homogeneous outer envelope. However, as discussed by \cite{Helled2022} and more recently by \cite{Fuentes_et_al2025}, the convective energy of this region is presumably not enough for stirring up the planet's core, at least for standard evolution calculations that adopt mixing-length theory. Double-diffusive convection provides a potentially different mechanism for transporting heat and heavy elements in regions where a stabilizing compositional gradient coexists with a destabilizing thermal stratification, as may occur in the cores of giant planets \citep{Leconte2012}. Current 3D simulations \citep[e.g.,][]{Tulekeyev2024} suggest that this process can organize into convective layers or staircases and potentially produce substantial mixing as these structures interact and merge. If efficient over evolutionary timescales, such mixing could redistribute the primordial composition gradients produced during formation and transform initially compact heavy-element distributions into partially mixed, spatially extended regions. 

Several prescriptions for double-diffusive mixing in the low-viscosity, high-thermal diffusivity regime relevant to astrophysical interiors have been proposed. These are based either on theoretical arguments \citep[e.g.,][]{Langer1983, Spruit2013} or on fits to three-dimensional numerical simulations at low diffusivities \citep[e.g.,][]{Wood2013,Moll2017}. The resulting scalings differ significantly from those derived in laboratory experiments of saltwater systems, which operate in a very different parameter regime \citep[e.g.,][]{Linden1978,Fernando1989}. Applying these prescriptions to interior profiles derived from formation models provides a way to quantify whether double-diffusive transport can substantially redistribute heavy elements, particularly at early times following accretion, when partial degeneracy allows thermal buoyancy to drive non-negligible mixing.

Double-diffusive transport have been incorporated into interior models of gas giants in several ways. For example, \cite{Leconte2012} built static structure models in which the heavy-element gradient in the interior is organized into a double-diffusive staircase, parameterizing the heat flux across the convective layers and diffusive interfaces combining mixing length theory and the numerical results of \cite{Rosenblum2011}. More recently, \cite{Knierim2024} treated double-diffusive transport by instantaneously mixing regions that are Ledoux stable but Schwarzwald unstable to convection, rather than applying a transport efficiency that depends on the local stability ratio. \cite{Muller2020} evolved Jupiter from formation motivated initial conditions including double-diffusive transport, finding that it had little effect on the resulting planetary structure. Their treatment, however, followed the prescription of \cite{Langer1983}, in which double-diffusive transport is represented by a small effective diffusion coefficient for composition, decoupled from the evolving thermal stratification. 

In this work, we incorporate the layered-convection prescription of \cite{Spruit2013} into the planetary evolution code \texttt{APPLE} \citep{Sur2024}, treating the transport of heat and composition simultaneously. Unlike previous treatments of double-diffusive transport, the efficiency of thermal and compositional mixing in this prescription is set locally and self-consistently by the thermal and compositional gradients, and evolves as the planet cools. This allows us to test whether the conclusion of \cite{Muller2020} persists under a more physically complete treatment. To our knowledge, this is the first application of such a prescription to post-formation interior models of giant planets.

In Section~\ref{sec:ddc}, we briefly review the double-diffusive instability, the regimes in which it can occur given the material properties of giant-planet interiors, and the layered-convection prescription adopted here. Section~\ref{sec:evolution} describes our interior models and main results. In Section~\ref{sec:energy}, we present an energetic argument for why double-diffusive mixing is inefficient in giant planets. We conclude in Section~\ref{sec:conclusions} with a summary and discussion.

\section{Double-Diffusive Convection}\label{sec:ddc}

Giant planets are expected to contain regions in which a destabilizing thermal stratification coexists with a stabilizing compositional stratification \citep{Stevenson1985}. Defining the thermal and compositional gradients,

\begin{equation}
\nabla \equiv \dfrac{d\ln T}{d\ln P},
\quad
\nabla_{\rm ad} \equiv
\left(\dfrac{\partial \ln T}{\partial \ln P}\right)_{S},
\quad
\nabla_\mu \equiv \dfrac{d\ln\mu}{d\ln P},
\end{equation}
together with the thermodynamic coefficients

\begin{equation}
\alpha_T \equiv
-\left(\dfrac{\partial \ln \rho}{\partial \ln T}\right)_{P,\mu},
\qquad
\alpha_\mu \equiv
\left(\dfrac{\partial \ln \rho}{\partial \ln \mu}\right)_{P,T},
\end{equation}
this translates to

\begin{equation}
0 < \nabla - \nabla_{\rm ad} < \dfrac{\alpha_\mu}{\alpha_T}\nabla_\mu,
\end{equation}
i.e., the fluid is unstable to overturning convection according to the Schwarzschild criterion but stable according to the Ledoux criterion \citep{Ledoux1947,Schwar1958}\footnote{In the stellar literature, this condition is often written in terms of the Ledoux gradient, $\nabla_{\rm L}=\nabla_{\rm ad}+(\alpha_\mu/\alpha_T)\nabla_\mu$, which reduces to $\nabla_{\rm ad}<\nabla<\nabla_{\rm L}$. For an ideal gas, $\alpha_T=\alpha_\mu=1$.}.

In this regime, the fluid is unstable to double-diffusive convection \citep[see review by][]{Garaud2021}. The origin of the instability is simple: heat diffuses significantly more rapidly than composition, so a displaced fluid element adjusts thermally before it can equilibrate compositionally, leading to oscillations of growing amplitude on a thermal timescale \citep{Kato1966}. The nonlinear outcome of the instability is not unique. One possibility is a weakly turbulent state--often referred to as homogeneous double-diffusive convection--in which transport is only modestly enhanced relative to pure diffusion. Alternatively, the system may organize into a staircase: a sequence of well-mixed convective layers separated by thin, stably stratified interfaces. Which of these states is achieved depends sensitively on the relative strength of the thermal and compositional gradients, as well as on the microscopic diffusivities. Introducing the stability ratio\footnote{There is some ambiguity in the literature, where some studies, e.g., \cite{Leconte2012}, refer to this parameter as the inverse density ratio, $R_\rho^{-1}$}

\begin{gather}
R_\rho = \dfrac{\alpha_\mu}{\alpha_T}\frac{\nabla_\mu}{\nabla - \nabla_{\rm ad}},
\end{gather}
double-diffusive instabilities are expected to arise within a finite range, 
\begin{align}
 1 < R_\rho < \dfrac{\mathrm{Pr} + 1}{\mathrm{Pr} + \tau}~,
\end{align}
\citep[e.g.,][]{Garaud2021}, where $\mathrm{Pr} = \nu/k_T$ is the Prandtl number, with $\nu$ the kinematic viscosity and $k_T$ the thermal diffusivity, and $\tau = D/k_T$, with $D$ the compositional (molecular) diffusivity. 

\begin{deluxetable}{lccccc}
\caption{Material properties (thermal diffusivity, kinematic viscosity, and compositional diffusivity) for giant-planet interiors under Mbar-pressure conditions. Numerical estimates for Jupiter are taken from \cite{French2012}, while those for Saturn are taken from \cite{Preising2023}. The compositional diffusivity is estimated from density functional molecular dynamics (DFT-MD) simulations by \cite{Wilson2015}, who found that for a heavy-element species $i$ in a H--He plasma at Mbar pressures, $D_i \sim D_{\mathrm{H}}(m_{\mathrm{H}}/m_i)^{1/2}$, where $m_{\mathrm{H}}$ and $m_i$ are the atomic masses of hydrogen and species $i$, respectively.\label{tab:giant_diffusivities}}
\tablewidth{\columnwidth}
\tablehead{
\colhead{Planet} &
\colhead{$k_T$} &
\colhead{$\nu$} &
\colhead{$D_{\rm H}$} &
$D_{\rm{Ice}}$ \\
\colhead{} &
\colhead{$(\mathrm{cm^2~s^{-1}})$} &
\colhead{$(\mathrm{cm^2~s^{-1}})$} &
\colhead{$(\mathrm{cm^2~s^{-1}})$} &
\colhead{$(\mathrm{cm^2~s^{-1}})$} 
}
\startdata
Jupiter & $2\times 10^{-1}$ & $3\times 10^{-3}$ & $4\times 10^{-3}$ & $1\times 10^{-3}$& \\
Saturn  & $6\times 10^{-2}$ & $4\times 10^{-3}$ & $3\times 10^{-3}$ & $7\times 10^{-4}$ 
\enddata
\end{deluxetable}

For giant planet conditions at Mbar pressures, thermal transport is dominated by degenerate electrons \citep{French2012}, while the microscopic transport of both momentum and composition is set by ion-ion interactions \citep{Wilson2015}. Representative values of the corresponding transport coefficients are given in Table~\ref{tab:giant_diffusivities}. In both Jupiter and Saturn, the thermal diffusivity greatly exceeds the viscosity and compositional diffusivity, implying $\mathrm{Pr} < 1$ and $\tau < 1$. The estimated values are $\mathrm{Pr} \approx 10^{-2}$ for Jupiter and $\mathrm{Pr} \approx 0.1$ for Saturn. Taking $m_i\approx 18$, appropriate for water-rich material gives $\tau \sim 5\times 10^{-3}$ for Jupiter and $\tau \sim 10^{-2}$ for Saturn. Taking these values, the upper bound for double-diffusive instability, $R_{\rho,\max}= (1+\mathrm{Pr})/(\tau+\mathrm{Pr})$, is approximately 67 for Jupiter and 10 for Saturn. Double-diffusive instabilities are therefore expected over a substantial range of stability ratios in both planets, although layer formation appears to require a more restricted range, $R_\rho \in [1, R_L]$, with $R_L$ close to one but not well constrained \citep{Mirouh2012,Wood2013,Fuentes2022,Tulekeyev2024,Fuentes2025}.

Since the three-dimensional nature of the nonlinear dynamics of layered convection cannot be represented explicitly in a one-dimensional evolutionary calculation, double-diffusive convection is parameterized by modeling the turbulent fluxes across layers as an enhanced diffusive process. Deriving prescriptions for this effective diffusivity is challenging because it generally requires numerical simulations of layered convection. Moreover, current hydrodynamical simulations show that once layers form, they tend to merge rapidly, ultimately leading to a single convection zone \citep[e.g.,][]{Tulekeyev2024,Fuentes2025}. As a result, it has not yet been possible to perform a comprehensive numerical study of a long-lived staircase in statistical equilibrium and consequently, existing prescriptions depend on an assumed layer thickness, $d$, whose value is not known a priori.

In the absence of a well-defined layer scale, it is natural to consider asymptotic regimes in which the transport becomes independent of $d$. \cite{Spruit2013} argued that this occurs when the convective layers are thin compared with the pressure scale height and the flow is turbulent, corresponding to large Reynolds and Péclet numbers. In this limit, they derive that the effective compositional and thermal diffusivities take the form

\begin{align}\label{eq:eff_diffusivities}
D_{\rm eff} \sim k_T\tau^{1/2} R_\rho^{-1},\quad k_{T,\rm eff} \sim k_T \tau^{-1/2} R^{-1}_{\rho},
\end{align}
so that $k_{T,\rm eff}/D_{\rm eff} = \tau^{-1}\gg 1$. The main assumptions underlying this scaling are that (i) the composition and heat fluxes across the well-mixed regions of the staircase (i.e., the convective layers) must match the diffusive fluxes at the adjacent stable interfaces, and (ii) only fluid parcels that are buoyantly unstable can be entrained by convective plumes, which limits the composition contrast carried across each well-mixed region to a fraction $1/R_\rho$ of the total contrast across a full double-diffusive step. Although both assumptions have been supported by numerical simulations in Cartesian geometry \citep{Zaussinger2013}, this prescription has not yet been incorporated in planetary evolution simulations. Spherical geometry should be irrelevant for this consideration (though still relevant for the overall energy budget considered later below). We therefore adopt this formulation in our evolutionary models to assess whether double-diffusive mixing can redistribute heavy material from the core into the overlying convective envelope and thereby influence the long-term cooling evolution of giant planets.

\section{Evolution Models}\label{sec:evolution}

To assess the long-term consequences of double-diffusive convection, it is necessary to evolve the planetary structure and the transport simultaneously, since the thermal stratification that drives the instability evolves together with the composition  profile. In this section, we first describe the post-formation interior structures that we use as initial conditions, then summarize how the double-diffusive transport prescription is implemented in the evolution code, and finally present our results.

\subsection{Initial Conditions from Formation Models}
\begin{figure*}
    \centering
    \includegraphics[width=\textwidth]{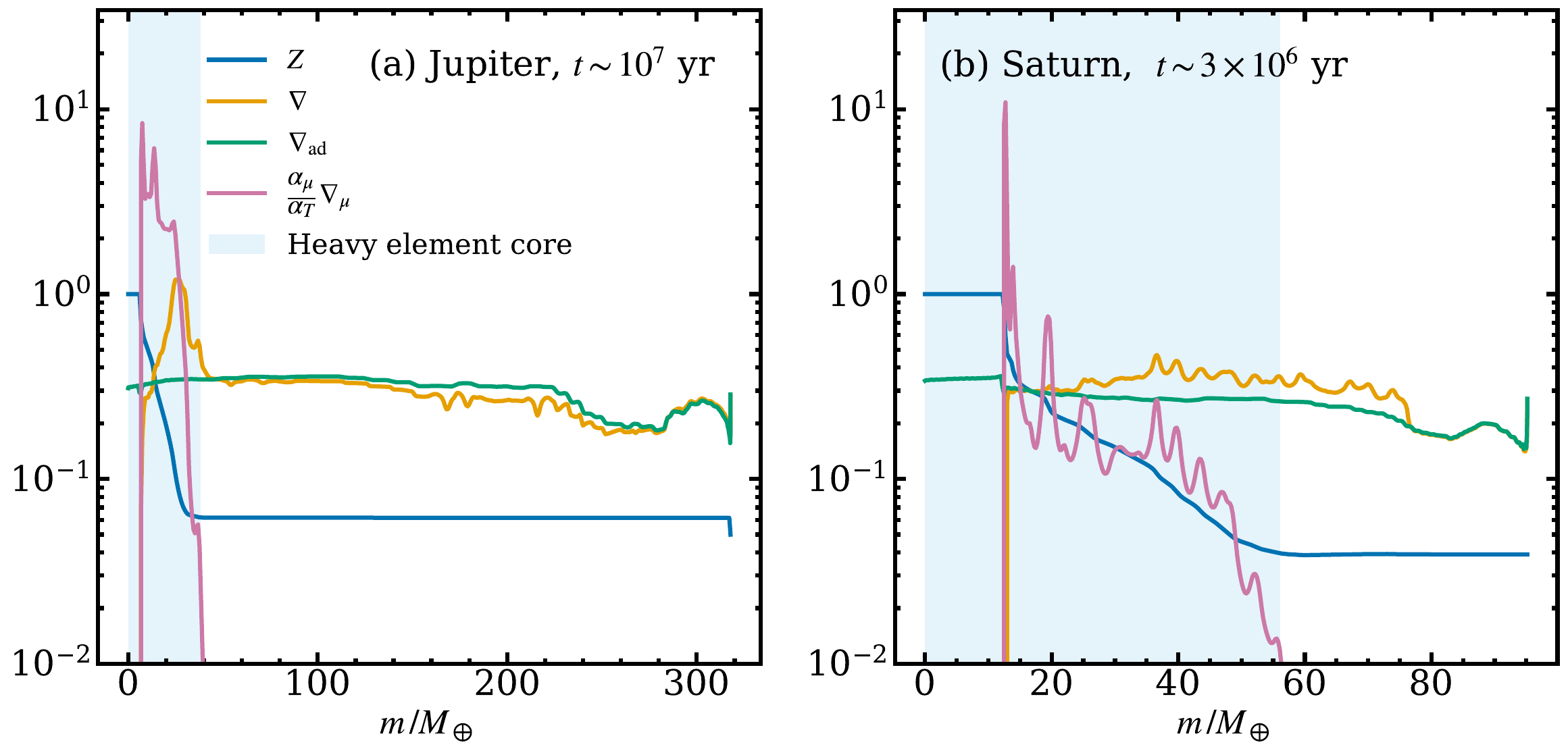}
    \caption{Internal profiles of the mass fraction of heavy elements, $Z$, and the relevant gradients for convective and double-diffusive mixing, $\nabla$, $\nabla_{\rm ad}$, and $(\alpha_\mu/\alpha_T)\nabla_\mu$. These profiles were taken from the formation models of Jupiter \citep{Stevenson2022}  and Saturn  \citep{Bodenheimer2025} at the end of accretion. The blue shaded region represents the extended heavy element cores for each planet.}
    \label{fig:initial_conditions}
\end{figure*}
 We adopt post-formation interior structures for Jupiter \citep{Stevenson2022} and Saturn \citep{Bodenheimer2025} as initial conditions. Both formation models follow the core accretion scenario with a one-dimensional Henyey code, in which accreted planetesimals ablate, break up, and dissolve in the growing H/He envelope instead of sinking to a central core. The planets form in situ, without orbital migration. Jupiter forms at 5.2 AU in a disk with initial solid surface density $\sigma = 10~\mathrm{g~cm^{-2}}$, whereas Saturn forms at 9.54 AU in a disk with $\sigma = 3~\mathrm{g~cm^{-2}}$. 

All planetesimals are assumed to have a single radius of $100~\mathrm{km}$, their composition is $50\%$ ice and $50\%$ rock in the Jupiter model, while $50\%$ ice, $35\%$ rock, and $15\%$ iron for the Saturn model. Planetesimal dynamics are not follow with N-body simulations. Instead, the solid accretion rates follows a statistical (Safronov) prescription that includes the gravitational enhancement factor of \cite{Greenzweig1992}. Once gas accretion becomes limited by the disk, the solid accretion rate is tied to the gas accretion rate through the expansion of the planet's feeding zone, and the gas accretion rates are taken from the three-dimensional simulations of disk--planet interactions \citep{Lissauer2009,DAngelo2021}. The Jupiter model accretes a total of $\approx 30~M_{\oplus}$ of heavy elements by the end of gas accretion ($t\approx 3.1~\mathrm{Myr}$), plus an additional $1~M_{\oplus}$ of planetesimals prescribed over the following 10 $\mathrm{Myr}$, for a final heavy element mass of $\approx 31~M_{\oplus}$. The Saturn model accretes $20~M_{\oplus}$ of heavy elements by the end of gas accretion ($t\approx 2.9~\mathrm{Myr}$).

Neither model includes double-diffusive mixing in the composition-gradient regions, which makes them suitable initial conditions for isolating its effect during the subsequent evolution. We refer the reader to \cite{Stevenson2022} and \cite{Bodenheimer2025} for further details.

These formation models lead naturally to fuzzy-core structures rather than sharp core boundaries (see Figure~\ref{fig:initial_conditions}). Outside the composition-gradient region, the envelope is convective, and the heavy elements delivered by planetesimals that dissolve there during the late stages of accretion are rapidly mixed, producing a uniform region where the mass fraction of heavy elements $Z$ is constant. The value of $Z$ in the convective envelope is set mainly during the disk-limited phase of gas accretion and depends on poorly constrained formation parameters; for example, in the Jupiter models of \citet{Stevenson2022}, reducing the width of the planet's feeding zone lowers the envelope $Z$ from $\approx 0.06$ to $\approx 0.02$. In addition, the Jupiter profile adopted here ($t \approx 10^7~\mathrm{yr}$) includes a period of solid accretion after gas accretion ends, which slightly increases the envelope $Z$, whereas the Saturn profile of \citet{Bodenheimer2025} ($t \approx 3\times 10^6~\mathrm{yr}$) corresponds to the end of gas accretion, with no subsequent solid accretion. The envelope metallicities of these models are therefore an outcome of the assumed formation parameters rather than being calibrated to atmospheric abundances. Since double-diffusive transport operates only within the composition-gradient region, the exact value of $Z$ in the convective envelope does not change our main conclusions.

We emphasize that our goal is to determine whether subsequent double-diffusive convection naturally transforms these post-formation structures into broader heavy-element distributions. The use of formation models is motivated not only by their physical basis, but also because they naturally produce density profiles with $R_\rho$ approaching unity over much of the fuzzy core. This is the regime in which layered convection is expected to develop. Artificially prescribed initial conditions need not lie in this regime and may therefore fail to develop double-diffusive layering.

\subsection{Evolution Code and Transport Prescription}

We evolve these models self-consistently for 4.56 Gyr using the evolution code \texttt{APPLE} \citep{Sur2024}, which solves the structure and transport equations on a Lagrangian mass grid using a Henyey scheme. The envelope hydrogen-helium equation of state is taken from \citet{Chabrier2021}, with heavy elements described by the AQUA EOS \citep{Haldemann2020}. The central ($Z=1$) region is treated as fully differentiated heavy material and assumed to be purely conductive. We do not include Helium rain since atmospheric helium abundances are not targeted in this study.

Regions unstable to overturning convection according to the Ledoux criterion are modeled using the standard mixing-length diffusivity $D_{\rm MLT} = v_{\rm MLT}l/3$, which rapidly homogenizes composition. In Ledoux-stable regions, we distinguish between purely diffusive evolution and double-diffusive transport. Where the stability ratio satisfies $1 < R_\rho < \tau^{-1/2}$, double-diffusive transport is parameterized using the effective compositional and thermal transport coefficients $(D_{\rm eff}, \kappa_{T,\rm eff})$ following Equation~\eqref{eq:eff_diffusivities}, representing transport by layered convection. Outside this interval, transport reduces to microscopic diffusion only, with $D_{\rm eff} = \kappa_{T,\rm eff} = 0$.
\begin{figure*}
    \centering    \includegraphics[width=\textwidth]{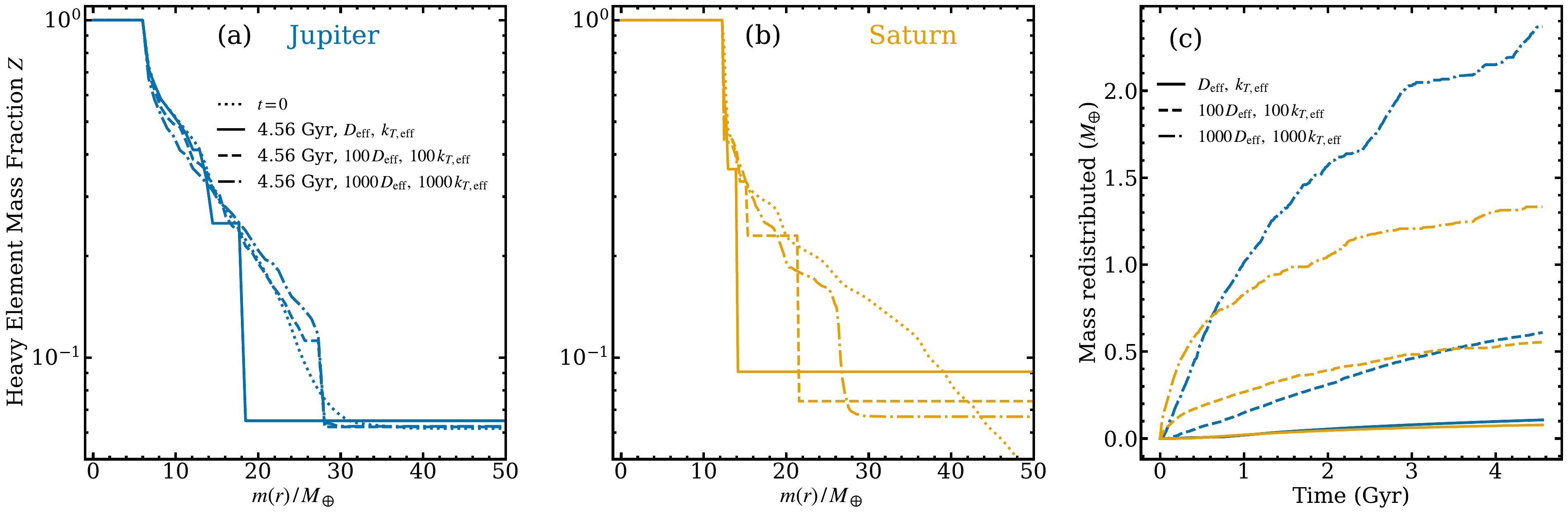}
    \caption{Heavy-element transport in Jupiter and Saturn. Initial post-formation heavy-element distributions adopted from the Jupiter model of \citet{Stevenson2022} and the Saturn model of \citet{Bodenheimer2025}, together with the distributions after 4.56 Gyr of evolution computed with \texttt{APPLE}. Panels (a) and (b) show results for Jupiter and Saturn, respectively. Different lines distinguish between the fiducial transport prescription and models in which both $D_{\rm eff}$ and $k_{T,\rm eff}$ are enhanced by factors of 100 and 1000. For better visualization of the heavy-element distribution in the fuzzy core, the horizontal axis shows the mass coordinate up to 50 $M_{\oplus}$. Panel (c): Cumulative mass of heavy elements transported from the core into the envelope by double-diffusive convection over the course of the evolution. Without double-diffusive transport, the composition profiles remain nearly unchanged from their initial ($t=0$) state after 4.56 Gyr of cooling.}
    \label{fig:results}
\end{figure*}

\subsection{Results}

Figure~\ref{fig:results} shows the evolution of the heavy-element distribution in Jupiter and Saturn. We consider three transport prescriptions: the fiducial model, using the baseline values of $D_{\rm eff}$ and $k_{T,\rm eff}$; a model in which both diffusivities are increased by a factor of 100; and a model in which they are increased by a factor of 1000. Panels (a) and (b) show that all models exhibit some inward displacement of the core-envelope transition relative to the initial profile. Since this shift is already present in the fiducial calculation, where double-diffusive transport is negligible, it is primarily the consequence of convective erosion from above. Increasing the effective diffusivities broadens the composition gradient, producing additional mixing within the fuzzy core but leaving the overall structure largely intact.

We emphasize that the step-like features visible in the evolved composition profiles should not be interpreted as resolved double-diffusive staircases. Such steps are a generic feature of one-dimensional evolution calculations, in which adjacent zones that cool at different rates can become locally unstable to convection according to the Ledoux criterion and are then instantaneously homogenized, without resolving the diffusive interfaces that would separate genuine double-diffusive layers \citep[see Section 3.2 of][]{Vazan2018}. In our models, double-diffusive staircases are not resolved explicitly. Their effect is represented only through the sub-grid effective diffusivities $D_{\rm eff}$ and $k_{T,\rm eff}$ of Equation~\eqref{eq:eff_diffusivities}.

Panel (c) quantifies the cumulative outward transport of heavy elements. In the fiducial calculation, only $\approx0.11\,M_{\oplus}$ is redistributed in Jupiter and $\approx0.07\,M_{\oplus}$ in Saturn over 4.56 Gyr. Increasing both diffusivities by a factor of 100 raises these values to only $\approx0.55\,M_{\oplus}$ and $\approx 0.61\,M_{\oplus}$, respectively, while the $\times 1000$ models redistribute $\approx 2\,M_{\oplus}$ in Jupiter and $\approx1.3\,M_{\oplus}$ in Saturn. 
Remarkably, increasing the effective diffusivities by up to three orders of magnitude produces only a modest increase in the amount of heavy material redistributed. This weak sensitivity demonstrates that our conclusions do not depend critically on the particular transport coefficients. Instead, the transport becomes self-limiting: enhanced thermal transport modifies the thermal stratification, reducing $R^{-1}_\rho$ and thereby suppressing both effective diffusivities through their nonlinear dependence on the stability ratio. This negative feedback prevents runaway mixing, so that even unrealistically large transport coefficients fail to produce substantial homogenization of the primordial composition gradient.

\section{An Energetic Constraint on Double-Diffusive Mixing}\label{sec:energy}

Our calculations demonstrate that even assuming persistent staircases exist throughout evolution, the associated large-scale transport of heavy elements by double-diffusive mixing in giant planets is remarkably inefficient, allowing primordial composition gradients to survive for Gyr timescales. Therefore, it is unlikely that an extended distribution of heavy material can result from double-diffusive mixing alone. 
This naturally raises the question of why self-consistent evolutionary models predict such weak mixing, whereas local hydrodynamical simulations often predict rapid layer mergers and efficient compositional transport. The answer lies in the physical assumptions adopted by the simulations.

Current simulations have been key in advancing our understanding of the nature of the instability \citep{Rosenblum2011}, conditions for layer formation \citep{Mirouh2012}, transport across diffusive interfaces \citep{Wood2013}, and the influence of additional physics such as rotation and magnetic fields \citep{Moll2017_rot, Sangui2022, Fuentes2024}. However, they remain limited by the use of the Boussinesq approximation \citep{Spiegel_Veronis_1960}. In this approximation, density variations across a fluid layer are assumed to be small (i.e., the flow is approximately incompressible), and perturbations from the reference density $\rho_0$ are related linearly to temperature and compositional fluctuations

\begin{align}
    \rho' = \rho_0 (\beta C' - \alpha T'),
    \end{align}
where $\alpha$ and $\beta$ are the thermodynamic coefficients of thermal expansion and compositional contraction, respectively.

To understand why mixing is so efficient in this limit, consider a fluid of thickness $H$ with linear temperature and compositional profiles,
\begin{align}
T(z)=\Delta T\left(1-\dfrac{z}{H}\right), \quad
C(z)=\Delta C\left(1-\dfrac{z}{H}\right),
\end{align}
where both temperature and heavy-element abundance increase with depth. We assume the fluid is stably stratified,
\begin{equation}
\frac{\beta\Delta C}{\alpha\Delta T}>1,
\end{equation}
so that the stabilizing compositional buoyancy exceeds the destabilizing thermal buoyancy.

The energy required to homogenize the compositional gradient across the layer (per unit area) is

\begin{equation}
E_{\rm mix} = \dfrac{\rho_0 \beta g H^3}{12}\left|\dfrac{dC}{dz}\right|,
\end{equation}
where we have used $\Delta C = H\left|dC/dz\right|/2$, which follows from integrating the linear profile.

The ratio of this mixing energy to the thermal energy available within the layer is then 

\begin{equation}
\dfrac{E_{\rm mix}}{\rho_0 c_P\Delta T\, H} = \dfrac{1}{6} \left(\dfrac{\beta\Delta C}{\alpha \Delta T}\right) \left(\dfrac{\alpha g H}{ c_P}\right) \sim \dfrac{\alpha g H}{c_P} \sim \dfrac{H}{H_T},
\label{eq:energyfrac}
\end{equation}
where $H_T=c_P/\alpha g$ is the thermal scale height and we have used $\beta \Delta C/\alpha \Delta T \sim 10$ for a layer unstable to double-diffusive convection. 

To assess how much of the convective energy budget is available to perform work, we use mixing length theory.  The thermal flux and convective speed are $F\sim \rho_0 c_P \delta T  v_{\rm conv} $ and $v_{\rm conv}^2\sim \alpha g H \delta T$, respectively, where $\delta T$ is a typical temperature fluctuation in the convective layer. The fraction of the convective heat flux that goes into kinetic energy is therefore
\begin{equation}
    \dfrac{F_{\rm KE}}{F}\sim \dfrac{\rho_0 v_{\rm conv}^3}{\rho_0 v_{\rm conv} c_P \delta T}\sim \dfrac{v_{\rm conv}^2}{c_P\delta T}  \sim \dfrac{\alpha g H}{c_P} \sim \dfrac{H}{H_T}. \label{eq:fluxfrac}
\end{equation}
Equations \eqref{eq:energyfrac}--\eqref{eq:fluxfrac} show that the energetic cost of mixing and the fraction of the convective heat flux available to perform mechanical work scale in the same way. In a Boussinesq fluid, where $H/H_T\ll1$ ($\sim 10^{-6}$ in water-based experiments, see \citealt{Zaussinger2019,Fuentes2020}), mixing is energetically cheap and occurs on short timescales. In giant planets, however, the stable region may span a substantial fraction of the planetary radius, $H\sim0.25$--$0.5R_p$, while $H_T\sim R_p$. Consequently, $H/H_T$ is of order unity, so the energy required to mix a stable compositional gradient can be comparable to the thermal energy available in the region of interest, i.e, mixing is no longer energetically cheap \citep[see also the discussion of core erosion in][]{Helled2022,Fuentes_et_al2025}. Simulations including many density scale heights are currently underway (Shu Zhang, in preparation), and will help further quantify the energy requirements for compositional mixing.

\section{Discussion}\label{sec:conclusions}

Our results show that double-diffusive convection, even when treated self-consistently throughout the planet's evolution, is unable to substantially broaden the primordial heavy-element distributions predicted by current formation models of Jupiter and Saturn. Our conclusions should be interpreted within the context of the adopted initial conditions. Specifically, we investigate whether composition gradients produced by current planet formation models are substantially modified by double-diffusive convection. Other initial conditions, such as those with an already dilute core \citep[e.g.,][]{Leconte2012}, may evolve differently. However, adopting such structures as initial conditions would address a different question, namely how those structures evolve once already established, rather than whether double-diffusive convection is capable of producing them.

If the fuzzy cores of gas giants cannot be explained by post-formation double-diffusive transport, the question becomes whether they are instead a direct product of formation itself, or whether some other process is responsible. Formation models naturally produce compositional gradients through the progressive dissolution and evaporation of accreted planetesimals within the growing H/He envelope \citep{Stevenson2022,Bodenheimer2025}, and the initial conditions adopted in this work are derived from precisely such models. However, the gradients they produce, while extended relative to a sharp core boundary, may still be too steep and too centrally concentrated to match current observational constraints without some degree of subsequent broadening.

One possibility is that large impacts during or shortly after the late stages of accretion could have delivered substantial heavy-element material at depths well above the primordial core boundary, effectively broadening the compositional gradient from the outside in (e.g., \citealt{Liu2019}). However, this scenario requires a head-on collision with a $\sim 10~M_{\oplus}$ impactor, and such events are expected to be rare according to N-body simulations \citep{Meier2025}. Moreover, recent hydrodynamic simulations suggest that, following a giant impact, heavy elements rapidly sediment and re-differentiate, making it challenging to produce a long-lived fuzzy core \citep{Sandnes2025}.

A second possibility is that standard formation models themselves underestimate the degree of early mixing. The treatment of planetesimal ablation and envelope interaction is necessarily simplified, and the actual distribution of dissolved material during accretion may be broader than current models predict. Similarly, the transition from the runaway gas accretion phase to a quasi-static cooling interior involves highly dynamical conditions that are not well captured by 1D evolutionary calculations.

A third possibility is that the sequence of events during formation is incorrect in the models we used. The standard assumption is that an embryo quickly forms from a planetesimal disk, followed by much slower accretion of planetesimals and gas, leading up to a ``runaway'' of the gas-rich envelope. Perhaps the differential orbital migrations of gas and solids are more complicated than this, with solids present both as large and small bodies (pebbles). Pebbles are expected to ablate and vaporize at larger radii within the envelope than planetesimals, depositing heavy elements farther from the center and producing more extended composition gradients \cite{Valletta2020}. In addition, pebbles drifting inward through the disk can sublimate, enriching the disk gas in heavy elements, which may then be accreted by the planet as vapor-rich gas \citep{Schneider2021}. Both processes could deliver heavy elements to the planet in a more extended way than the planetesimal accretion assumed in the formation models adopted here.

Taken together, these possibilities suggest that the origin of fuzzy cores in Jupiter and Saturn remains an open problem. Our results indicate that double-diffusive convection, at least as currently understood and parameterized, is unlikely to provide the required post-formation mixing. Instead, the observed compositional structure may primarily reflect processes operating during planet formation itself, or other transport mechanisms that have yet to be explored. Improving our understanding of these possibilities will require combining increasingly realistic simulations of planet formation and giant impacts with evolutionary models capable of following the long-term thermal and compositional evolution of giant planets. Such efforts will be essential for connecting the interior structures revealed by Juno and Cassini to the physical processes that shaped the formation and evolution of Jupiter and Saturn.

\begin{acknowledgements}
We thank the referee for a thorough and insightful report that improved the paper, and Simon Müller and Ravit Helled for addressing our questions regarding the physics implemented in their formation models. J.R.F. is supported by the Sherman Fairchild Postdoctoral (Burke) Fellowship and the Presidential Fellowship at Caltech, as well as NASA Solar System Workings grant 80NSSC24K0927. 
\end{acknowledgements}




\bibliography{references}{}

@ARTICLE{Stevenson2022,
       author = {{Stevenson}, David J. and {Bodenheimer}, Peter and {Lissauer}, Jack J. and {D'Angelo}, Gennaro},
        title = "{Mixing of Condensable Constituents with H-He during the Formation and Evolution of Jupiter}",
      journal = {PSJ},
         year = 2022,
        month = apr,
       volume = {3},
       number = {4},
          eid = {74},
        pages = {74},
          doi = {10.3847/PSJ/ac5c44},
archivePrefix = {arXiv},
       eprint = {2202.09476},
 primaryClass = {astro-ph.EP},
       adsurl = {https://ui.adsabs.harvard.edu/abs/2022PSJ.....3...74S}
}

@ARTICLE{Valletta2020,
       author = {{Valletta}, Claudio and {Helled}, Ravit},
        title = "{Giant Planet Formation Models with a Self-consistent Treatment of the Heavy Elements}",
      journal = {\apj},
         year = 2020,
        month = sep,
       volume = {900},
       number = {2},
          eid = {133},
        pages = {133},
          doi = {10.3847/1538-4357/aba904},
archivePrefix = {arXiv},
       eprint = {2007.13577},
 primaryClass = {astro-ph.EP},
       adsurl = {https://ui.adsabs.harvard.edu/abs/2020ApJ...900..133V}
}

@ARTICLE{Greenzweig1992,
       author = {{Greenzweig}, Yuval and {Lissauer}, Jack J.},
        title = "{Accretion rates of protoplanets  II. Gaussian distributions of planetesimal velocities}",
      journal = {\icarus},
         year = 1992,
        month = dec,
       volume = {100},
       number = {2},
        pages = {440-463},
          doi = {10.1016/0019-1035(92)90110-S},
       adsurl = {https://ui.adsabs.harvard.edu/abs/1992Icar..100..440G}
}

@ARTICLE{DAngelo2021,
       author = {{D'Angelo}, Gennaro and {Weidenschilling}, Stuart J. and {Lissauer}, Jack J. and {Bodenheimer}, Peter},
        title = "{Growth of Jupiter: Formation in disks of gas and solids and evolution to the present epoch}",
      journal = {\icarus},
         year = 2021,
        month = feb,
       volume = {355},
          eid = {114087},
        pages = {114087},
          doi = {10.1016/j.icarus.2020.114087},
archivePrefix = {arXiv},
       eprint = {2009.05575},
 primaryClass = {astro-ph.EP},
       adsurl = {https://ui.adsabs.harvard.edu/abs/2021Icar..35514087D}
}

@ARTICLE{Lissauer2009,
       author = {{Lissauer}, Jack J. and {Hubickyj}, Olenka and {D'Angelo}, Gennaro and {Bodenheimer}, Peter},
        title = "{Models of Jupiter's growth incorporating thermal and hydrodynamic constraints}",
      journal = {\icarus},
         year = 2009,
        month = feb,
       volume = {199},
       number = {2},
        pages = {338-350},
          doi = {10.1016/j.icarus.2008.10.004},
archivePrefix = {arXiv},
       eprint = {0810.5186},
 primaryClass = {astro-ph},
       adsurl = {https://ui.adsabs.harvard.edu/abs/2009Icar..199..338L}
}

@ARTICLE{Schneider2021,
       author = {{Schneider}, Aaron David and {Bitsch}, Bertram},
        title = "{How drifting and evaporating pebbles shape giant planets. I. Heavy element content and atmospheric C/O}",
      journal = {\aap},
         year = 2021,
        month = oct,
       volume = {654},
          eid = {A71},
        pages = {A71},
          doi = {10.1051/0004-6361/202039640},
archivePrefix = {arXiv},
       eprint = {2105.13267},
 primaryClass = {astro-ph.EP},
       adsurl = {https://ui.adsabs.harvard.edu/abs/2021A&A...654A..71S}
}

@ARTICLE{Moll2017,
  title     = "Double-diffusive erosion of the core of Jupiter",
  author    = "Moll, R and Garaud, P and Mankovich, C and Fortney, J J",
  journal   = "ApJ",
  publisher = "American Astronomical Society",
  volume    =  849,
  number    =  1,
  pages     = "24",
  month     =  oct,
  year      =  2017,
  copyright = "http://iopscience.iop.org/page/copyright"
}

@ARTICLE{Vazan2018,
       author = {{Vazan}, Allona and {Helled}, Ravit and {Guillot}, Tristan},
        title = "{Jupiter's evolution with primordial composition gradients}",
      journal = {\aap},
         year = 2018,
        month = feb,
       volume = {610},
          eid = {L14},
        pages = {L14},
          doi = {10.1051/0004-6361/201732522},
archivePrefix = {arXiv},
       eprint = {1801.08149},
 primaryClass = {astro-ph.EP},
       adsurl = {https://ui.adsabs.harvard.edu/abs/2018A&A...610L..14V}
}

@ARTICLE{Knierim2024,
       author = {{Knierim}, Henrik and {Helled}, Ravit},
        title = "{Convective Mixing in Gas Giant Planets with Primordial Composition Gradients}",
      journal = {\apj},
         year = 2024,
        month = dec,
       volume = {977},
       number = {2},
          eid = {227},
        pages = {227},
          doi = {10.3847/1538-4357/ad8dd0},
archivePrefix = {arXiv},
       eprint = {2407.09341},
 primaryClass = {astro-ph.EP},
       adsurl = {https://ui.adsabs.harvard.edu/abs/2024ApJ...977..227K}
}

@ARTICLE{Militzer2024,
       author = {{Militzer}, Burkhard and {Hubbard}, William B.},
        title = "{Study of Jupiter's interior: Comparison of 2, 3, 4, 5, and 6 layer models}",
      journal = {Icar},
         year = 2024,
        month = mar,
       volume = {411},
          eid = {115955},
        pages = {115955},
          doi = {10.1016/j.icarus.2024.115955},
archivePrefix = {arXiv},
       eprint = {2401.12166},
 primaryClass = {astro-ph.EP},
       adsurl = {https://ui.adsabs.harvard.edu/abs/2024Icar..41115955M}
}

@article{Langer1983,
  author  = {Langer, N. and Fricke, K. J. and Sugimoto, D.},
  title   = {Semiconvective diffusion and energy transport},
  journal = {A\&A},
  year    = {1983},
  volume  = {126},
  pages   = {207--208},
}

@ARTICLE{Meier2025,
       author = {{Meier}, Thomas and {Reinhardt}, Christian and {Shibata}, Sho and {M{\"u}ller}, Simon and {Stadel}, Joachim and {Helled}, Ravit},
        title = "{On the Origin of Jupiter's Fuzzy Core: Constraints from N-body, Impact, and Evolution Simulations}",
      journal = {\apj},
         year = 2025,
        month = jul,
       volume = {988},
       number = {1},
          eid = {7},
        pages = {7},
          doi = {10.3847/1538-4357/addbe6},
archivePrefix = {arXiv},
       eprint = {2503.23997},
 primaryClass = {astro-ph.EP},
       adsurl = {https://ui.adsabs.harvard.edu/abs/2025ApJ...988....7M}
}

@ARTICLE{Liu2019,
  title     = "The formation of Jupiter's diluted core by a giant impact",
  author    = "Liu, Shang-Fei and Hori, Yasunori and M{\"u}ller, Simon and
               Zheng, Xiaochen and Helled, Ravit and Lin, Doug and Isella,
               Andrea",
  journal   = "Nature",
  publisher = "Springer Science and Business Media LLC",
  volume    =  572,
  number    =  7769,
  pages     = "355--357",
  month     =  aug,
  year      =  2019,
  language  = "en"
}

@ARTICLE{Sandnes2025,
       author = {{Sandnes}, T.~D. and {Eke}, V.~R. and {Kegerreis}, J.~A. and {Massey}, R.~J. and {Teodoro}, L.~F.~A.},
        title = "{No dilute core produced in simulations of giant impacts on to Jupiter}",
      journal = {\mnras},
         year = 2025,
        month = sep,
       volume = {542},
       number = {2},
        pages = {947-959},
          doi = {10.1093/mnras/staf1105},
archivePrefix = {arXiv},
       eprint = {2412.06094},
 primaryClass = {astro-ph.EP},
       adsurl = {https://ui.adsabs.harvard.edu/abs/2025MNRAS.542..947S}
}

@ARTICLE{Mankovich2021,
       author = {{Mankovich}, Christopher R. and {Fuller}, Jim},
        title = "{A diffuse core in Saturn revealed by ring seismology}",
      journal = {NatAs},
         year = 2021,
        month = aug,
       volume = {5},
        pages = {1103-1109},
          doi = {10.1038/s41550-021-01448-3},
       adsurl = {https://ui.adsabs.harvard.edu/abs/2021NatAs...5.1103M}
}

@article{Wahl2017,
author = {Wahl, S. M. and Hubbard, W. B. and Militzer, B. and Guillot, T. and Miguel, Y. and Movshovitz, N. and Kaspi, Y. and Helled, R. and Reese, D. and Galanti, E. and Levin, S. and Connerney, J. E. and Bolton, S. J.},
title = {Comparing Jupiter interior structure models to Juno gravity measurements and the role of a dilute core},
journal = {Geophysical Research Letters},
volume = {44},
number = {10},
pages = {4649-4659},
doi = {https://doi.org/10.1002/2017GL073160},
url = {https://agupubs.onlinelibrary.wiley.com/doi/abs/10.1002/2017GL073160},
eprint = {https://agupubs.onlinelibrary.wiley.com/doi/pdf/10.1002/2017GL073160},
year = {2017}
}

@ARTICLE{Garaud2021,
       author = {{Garaud}, Pascale},
        title = "{Double-diffusive processes in stellar astrophysics}",
      journal = {arXiv e-prints},
         year = 2021,
        month = mar,
          eid = {arXiv:2103.08072},
        pages = {arXiv:2103.08072},
          doi = {10.48550/arXiv.2103.08072},
archivePrefix = {arXiv},
       eprint = {2103.08072},
 primaryClass = {astro-ph.SR},
       adsurl = {https://ui.adsabs.harvard.edu/abs/2021arXiv210308072G}
}

@ARTICLE{Schwar1958,
   author = {{Schwarzschild}, M. and {H{\"a}rm}, R.},
    title = "{Evolution of Very Massive Stars.}",
  journal = {ApJ},
     year = 1958,
    month = sep,
   volume = 128,
    pages = {348},
      doi = {10.1086/146548},
   adsurl = {https://ui.adsabs.harvard.edu/abs/1958ApJ...128..348S}
}

@ARTICLE{Preising2023,
       author = {{Preising}, Martin and {French}, Martin and {Mankovich}, Christopher and {Soubiran}, Fran{\c{c}}ois and {Redmer}, Ronald},
        title = "{Material Properties of Saturn's Interior from Ab Initio Simulations}",
      journal = {\apjs},
         year = 2023,
        month = dec,
       volume = {269},
       number = {2},
          eid = {47},
        pages = {47},
          doi = {10.3847/1538-4365/ad0293},
       adsurl = {https://ui.adsabs.harvard.edu/abs/2023ApJS..269...47P}
}

@ARTICLE{Stevenson1985,
       author = {{Stevenson}, D.~J.},
        title = "{Cosmochemistry and structure of the giant planets and their satellites}",
      journal = {Icar},
         year = 1985,
        month = apr,
       volume = {62},
       number = {1},
        pages = {4-15},
          doi = {10.1016/0019-1035(85)90168-X},
       adsurl = {https://ui.adsabs.harvard.edu/abs/1985Icar...62....4S}
}

@ARTICLE{Ledoux1947,
   author = {{Ledoux}, P.},
    title = "{Stellar Models with Convection and with Discontinuity of the Mean Molecular Weight}",
  journal = {ApJ},
     year = 1947,
    month = mar,
   volume = 105,
    pages = {305},
      doi = {10.1086/144905},
   adsurl = {https://ui.adsabs.harvard.edu/abs/1947ApJ...105..305L}
}

@ARTICLE{Sangui2022,
       author = {{Sanghi}, A. and {Fraser}, A.~E. and {Tian}, E.~W. and {Garaud}, P.},
        title = "{Magnetized Oscillatory Double-diffusive Convection}",
      journal = {\apj},
         year = 2022,
        month = aug,
       volume = {935},
       number = {1},
          eid = {33},
        pages = {33},
          doi = {10.3847/1538-4357/ac73ed},
archivePrefix = {arXiv},
       eprint = {2205.02251},
 primaryClass = {astro-ph.SR},
       adsurl = {https://ui.adsabs.harvard.edu/abs/2022ApJ...935...33S}
}

@ARTICLE{Moll2017_rot,
       author = {{Moll}, Ryan and {Garaud}, Pascale},
        title = "{The Effect of Rotation on Oscillatory Double-diffusive Convection (Semiconvection)}",
      journal = {ApJ},
         year = 2017,
        month = jan,
       volume = {834},
       number = {1},
          eid = {44},
        pages = {44},
          doi = {10.3847/1538-4357/834/1/44},
archivePrefix = {arXiv},
       eprint = {1610.03940},
 primaryClass = {astro-ph.SR},
       adsurl = {https://ui.adsabs.harvard.edu/abs/2017ApJ...834...44M}
}

@ARTICLE{Fuentes2024,
       author = {{Fuentes}, J.~R. and {Hindman}, Bradley W. and {Fraser}, Adrian E. and {Anders}, Evan H.},
        title = "{Evolution of Semiconvective Staircases in Rotating Flows: Consequences for Fuzzy Cores in Giant Planets}",
      journal = {\apjl},
         year = 2024,
        month = nov,
       volume = {975},
       number = {1},
          eid = {L1},
        pages = {L1},
          doi = {10.3847/2041-8213/ad84dc},
archivePrefix = {arXiv},
       eprint = {2408.10833},
 primaryClass = {astro-ph.EP},
       adsurl = {https://ui.adsabs.harvard.edu/abs/2024ApJ...975L...1F}
}

@ARTICLE{Rosenblum2011,
   author = {{Rosenblum}, E. and {Garaud}, P. and {Traxler}, A. and {Stellmach}, S.
	},
    title = "{Turbulent Mixing and Layer Formation in Double-diffusive Convection: Three-dimensional Numerical Simulations and Theory}",
  journal = {ApJ},
     year = 2011,
    month = apr,
   volume = 731,
      eid = {66},
    pages = {66},
      doi = {10.1088/0004-637X/731/1/66},
   adsurl = {https://ui.adsabs.harvard.edu/abs/2011ApJ...731...66R}
}

@ARTICLE{Tulekeyev2024,
       author = {{Tulekeyev}, A. and {Garaud}, P. and {Idini}, B. and {Fortney}, J.~J.},
        title = "{Constraints on the Long-term Existence of Dilute Cores in Giant Planets}",
      journal = {PSJ},
         year = 2024,
        month = aug,
       volume = {5},
       number = {8},
          eid = {190},
        pages = {190},
          doi = {10.3847/PSJ/ad6571},
archivePrefix = {arXiv},
       eprint = {2405.06790},
 primaryClass = {astro-ph.EP},
       adsurl = {https://ui.adsabs.harvard.edu/abs/2024PSJ.....5..190T}
}

@ARTICLE{Fuentes2022,
       author = {{Fuentes}, J.~R. and {Cumming}, A. and {Anders}, E.~H.},
        title = "{Layer formation in a stably stratified fluid cooled from above: Towards an analog for Jupiter and other gas giants}",
      journal = {PhRvF},
         year = 2022,
        month = dec,
       volume = {7},
       number = {12},
          eid = {124501},
        pages = {124501},
          doi = {10.1103/PhysRevFluids.7.124501},
       adsurl = {https://ui.adsabs.harvard.edu/abs/2022PhRvF...7l4501F}
}

@article{Helled2022,
title = {Revelations on Jupiter's formation, evolution and interior: Challenges from Juno results},
journal = {Icarus},
volume = {378},
pages = {114937},
year = {2022},
issn = {0019-1035},
doi = {https://doi.org/10.1016/j.icarus.2022.114937},
url = {https://www.sciencedirect.com/science/article/pii/S0019103522000586},
author = {Ravit Helled and David J. Stevenson and Jonathan I. Lunine and Scott J. Bolton and Nadine Nettelmann and Sushil Atreya and Tristan Guillot and Burkhard Militzer and Yamila Miguel and William B. Hubbard}
}

@ARTICLE{Fuentes2020,
       author = {{Fuentes}, J.~R. and {Cumming}, A.},
        title = "{Penetration of a cooling convective layer into a stably-stratified composition gradient: Entrainment at low Prandtl number}",
      journal = {Phys. Rev. Fluids},
         year = 2020,
        month = dec,
       volume = {5},
       number = {12},
          eid = {124501},
        pages = {124501},
          doi = {10.1103/PhysRevFluids.5.124501},
       adsurl = {https://ui.adsabs.harvard.edu/abs/2020PhRvF...5l4501F}
}

@ARTICLE{Zaussinger2019,
       author = {{Zaussinger}, Florian and {Kupka}, Friedrich},
        title = "{Layer formation in double-diffusive convection over resting and moving heated plates}",
      journal = {ThCFD},
         year = "2019",
        month = "Aug",
       volume = {33},
       number = {3-4},
        pages = {383-409},
          doi = {10.1007/s00162-019-00499-7},
       adsurl = {https://ui.adsabs.harvard.edu/abs/2019ThCFD..33..383Z}
}

@ARTICLE{Fuentes_et_al2025,
       author = {{Fuentes}, J.~R. and {Mankovich}, Christopher R. and {Sur}, Ankan},
        title = "{An Energy Perspective of Core Erosion in Gas Giant Planets}",
      journal = {\apjl},
         year = 2025,
        month = aug,
       volume = {988},
       number = {2},
          eid = {L49},
        pages = {L49},
          doi = {10.3847/2041-8213/adef0a},
archivePrefix = {arXiv},
       eprint = {2507.05109},
 primaryClass = {astro-ph.EP},
       adsurl = {https://ui.adsabs.harvard.edu/abs/2025ApJ...988L..49F}
}

@ARTICLE{Fuentes2025,
       author = {{Fuentes}, J.~R.},
        title = "{3D Simulations of Semiconvection in Spheres: Turbulent Mixing and Layer Formation}",
      journal = {\apj},
         year = 2025,
        month = mar,
       volume = {982},
       number = {1},
          eid = {44},
        pages = {44},
          doi = {10.3847/1538-4357/adb8ec},
archivePrefix = {arXiv},
       eprint = {2502.15111},
 primaryClass = {astro-ph.SR},
       adsurl = {https://ui.adsabs.harvard.edu/abs/2025ApJ...982...44F}
}

@ARTICLE{Zaussinger2013,
       author = {{Zaussinger}, F. and {Spruit}, H.~C.},
        title = "{Semiconvection: numerical simulations}",
      journal = {\aap},
         year = 2013,
        month = jun,
       volume = {554},
          eid = {A119},
        pages = {A119},
          doi = {10.1051/0004-6361/201220573},
archivePrefix = {arXiv},
       eprint = {1303.4522},
 primaryClass = {astro-ph.SR},
       adsurl = {https://ui.adsabs.harvard.edu/abs/2013A&A...554A.119Z}
}

@ARTICLE{Linden1978,
       author = {{Linden}, P.~F. and {Shirtcliffe}, T.~G.~L.},
        title = "{The diffusive interface in double-diffusive convection}",
      journal = {JFM},
         year = 1978,
        month = jan,
       volume = {87},
        pages = {417-432},
          doi = {10.1017/S002211207800169X},
       adsurl = {https://ui.adsabs.harvard.edu/abs/1978JFM....87..417L}
}

@ARTICLE{Fernando1989,
       author = {{Fernando}, H.~J.~S.},
        title = "{Buoyancy transfer across a diffusive interface}",
      journal = {JFM},
         year = 1989,
        month = jan,
       volume = {209},
        pages = {1-34},
          doi = {10.1017/S0022112089003010},
       adsurl = {https://ui.adsabs.harvard.edu/abs/1989JFM...209....1F}
}

@ARTICLE{Spruit2013,
       author = {{Spruit}, H.~C.},
        title = "{Semiconvection: theory}",
      journal = {\aap},
         year = 2013,
        month = apr,
       volume = {552},
          eid = {A76},
        pages = {A76},
          doi = {10.1051/0004-6361/201220575},
archivePrefix = {arXiv},
       eprint = {1302.4005},
 primaryClass = {astro-ph.SR},
       adsurl = {https://ui.adsabs.harvard.edu/abs/2013A&A...552A..76S}
}

@article{Leconte2012,
	author = {{Leconte, J.} and {Chabrier, G.}},
	title = {A new vision of giant planet interiors:  Impact of double diffusive convection},
	DOI= "10.1051/0004-6361/201117595",
	url= "https://doi.org/10.1051/0004-6361/201117595",
	journal = {A\&A},
	year = 2012,
	volume = 540,
	pages = "A20",
	month = "",
}

@ARTICLE{Tejada2025,
       author = {{Tejada Arevalo}, Roberto and {Sur}, Ankan and {Su}, Yubo and {Burrows}, Adam},
        title = "{Jupiter Evolutionary Models Incorporating Stably Stratified Regions}",
      journal = {ApJ},
         year = 2025,
        month = feb,
       volume = {979},
       number = {2},
          eid = {243},
        pages = {243},
          doi = {10.3847/1538-4357/ada030},
archivePrefix = {arXiv},
       eprint = {2410.12899},
 primaryClass = {astro-ph.EP},
       adsurl = {https://ui.adsabs.harvard.edu/abs/2025ApJ...979..243T}
}

@ARTICLE{Sur2025,
       author = {{Sur}, Ankan and {Tejada Arevalo}, Roberto and {Su}, Yubo and {Burrows}, Adam},
        title = "{Simultaneous Evolutionary Fits for Jupiter and Saturn Incorporating Fuzzy Cores}",
      journal = {\apjl},
         year = 2025,
        month = feb,
       volume = {980},
       number = {1},
          eid = {L5},
        pages = {L5},
          doi = {10.3847/2041-8213/adad62},
archivePrefix = {arXiv},
       eprint = {2412.17127},
 primaryClass = {astro-ph.EP},
       adsurl = {https://ui.adsabs.harvard.edu/abs/2025ApJ...980L...5S}
}

@ARTICLE{Knierim2026,
       author = {{Knierim}, H. and {Batygin}, K. and {Helled}, R. and {Morf}, L. and {Adams}, F.~C.},
        title = "{Further constraints on Jupiter's primordial structure}",
      journal = {\aap},
         year = 2026,
        month = jan,
       volume = {706},
          eid = {A51},
        pages = {A51},
          doi = {10.1051/0004-6361/202556984},
archivePrefix = {arXiv},
       eprint = {2512.03961},
 primaryClass = {astro-ph.EP},
       adsurl = {https://ui.adsabs.harvard.edu/abs/2026A&A...706A..51K}
}

@ARTICLE{Muller2020,
       author = {{M{\"u}ller}, Simon and {Helled}, Ravit and {Cumming}, Andrew},
        title = "{The challenge of forming a fuzzy core in Jupiter}",
      journal = {A\&A.},
         year = 2020,
        month = jun,
       volume = {638},
          eid = {A121},
        pages = {A121},
          doi = {10.1051/0004-6361/201937376},
       adsurl = {https://ui.adsabs.harvard.edu/abs/2020A&A...638A.121M}
}

@ARTICLE{Wilson2015,
       author = {{Wilson}, Hugh F.},
        title = "{Diffusivity of heavy elements in Jupiter and Saturn}",
      journal = {Icar},
         year = 2015,
        month = apr,
       volume = {250},
        pages = {400-404},
          doi = {10.1016/j.icarus.2014.11.031},
       adsurl = {https://ui.adsabs.harvard.edu/abs/2015Icar..250..400W}
}

@ARTICLE{French2012,
       author = {{French}, Martin and {Becker}, Andreas and {Lorenzen}, Winfried and
         {et al.}},
        title = "{Ab Initio Simulations for Material Properties along the Jupiter Adiabat}",
      journal = {ApJS},
         year = 2012,
        month = sep,
       volume = {202},
       number = {1},
          eid = {5},
        pages = {5},
          doi = {10.1088/0067-0049/202/1/5},
       adsurl = {https://ui.adsabs.harvard.edu/abs/2012ApJS..202....5F}
}

@ARTICLE{Mirouh2012,
       author = {{Mirouh}, G.~M. and {Garaud}, P. and {Stellmach}, S. and {Traxler}, A.~L. and {Wood}, T.~S.},
        title = "{A New Model for Mixing by Double-diffusive Convection (Semi-convection). I. The Conditions for Layer Formation}",
      journal = {ApJ},
         year = 2012,
        month = may,
       volume = {750},
       number = {1},
          eid = {61},
        pages = {61},
          doi = {10.1088/0004-637X/750/1/61},
archivePrefix = {arXiv},
       eprint = {1112.4819},
 primaryClass = {astro-ph.SR},
       adsurl = {https://ui.adsabs.harvard.edu/abs/2012ApJ...750...61M}
}

@ARTICLE{Wood2013,
       author = {{Wood}, T.~S. and {Garaud}, P. and {Stellmach}, S.},
        title = "{A New Model for Mixing by Double-diffusive Convection (Semi-convection). II. The Transport of Heat and Composition through Layers}",
      journal = {ApJ},
         year = 2013,
        month = may,
       volume = {768},
       number = {2},
          eid = {157},
        pages = {157},
          doi = {10.1088/0004-637X/768/2/157},
archivePrefix = {arXiv},
       eprint = {1212.1218},
 primaryClass = {astro-ph.SR},
       adsurl = {https://ui.adsabs.harvard.edu/abs/2013ApJ...768..157W}
}

@ARTICLE{Bodenheimer2025,
       author = {{Bodenheimer}, Peter and {Stevenson}, David J. and {Lissauer}, Jack J. and {D'Angelo}, Gennaro},
        title = "{Formation and Evolution Simulations of Saturn, Including Composition Gradients and Helium Immiscibility}",
      journal = {PSJ},
         year = 2025,
        month = jun,
       volume = {6},
       number = {6},
          eid = {143},
        pages = {143},
          doi = {10.3847/PSJ/add0b0},
archivePrefix = {arXiv},
       eprint = {2504.17227},
 primaryClass = {astro-ph.EP},
       adsurl = {https://ui.adsabs.harvard.edu/abs/2025PSJ.....6..143B}
}

@ARTICLE{Kato1966,
       author = {{Kato}, S.},
        title = "{Overstable Convection in a Medium Stratified in Mean Molecular Weight}",
      journal = {PASJ},
         year = 1966,
        month = jan,
       volume = {18},
        pages = {374},
       adsurl = {https://ui.adsabs.harvard.edu/abs/1966PASJ...18..374K}
}

@ARTICLE{Spiegel_Veronis_1960,
       author = {{Spiegel}, E.~A. and {Veronis}, G.},
        title = "{On the Boussinesq Approximation for a Compressible Fluid.}",
      journal = {ApJ},
         year = 1960,
        month = mar,
       volume = {131},
        pages = {442},
          doi = {10.1086/146849},
       adsurl = {https://ui.adsabs.harvard.edu/abs/1960ApJ...131..442S}
}

@article{Sur2024,
    title = {{APPLE: An Evolution Code for Modeling Giant Planets}},
    year = {2024},
    journal = {The Astrophysical Journal},
    author = {Sur, Ankan and Su, Yubo and Arevalo, Roberto Tejada and Chen, Yi-Xian and Burrows, Adam},
    number = {1},
    pages = {104},
    volume = {971},
    publisher = {IOP Publishing},
    url = {http://arxiv.org/abs/2404.14483},
    doi = {10.3847/1538-4357/ad57c3},
    issn = {1538-4357},
    arxivId = {2404.14483}
}

@ARTICLE{Sur2025b,
       author = {{Sur}, Ankan and {Burrows}, Adam and {Arevalo}, Roberto Tejada and {Su}, Yubo},
        title = "{The Evolution of Jupiter and Saturn as a Function of the R$_{{\ensuremath{\rho}}}$ Parameter}",
      journal = {\apj},
         year = 2025,
        month = dec,
       volume = {994},
       number = {2},
          eid = {186},
        pages = {186},
          doi = {10.3847/1538-4357/ae16a3},
archivePrefix = {arXiv},
       eprint = {2506.19041},
 primaryClass = {astro-ph.EP},
       adsurl = {https://ui.adsabs.harvard.edu/abs/2025ApJ...994..186S}
}

@article{Chabrier2021,
    title = {{A New Equation of State for Dense Hydrogen-Helium Mixtures. II. Taking into Account Hydrogen-Helium Interactions}},
    year = {2021},
    journal = {ApJ},
    author = {Chabrier, Gilles and Debras, Florian},
    number = {4},
    pages = {6pp},
    volume = {917},
    url = {https://doi.org/10.3847/1538-4357/abfc48},
    doi = {10.3847/1538-4357/abfc48}
}

@article{Haldemann2020,
    title = {{AQUA: a collection of H 2 O equations of state for planetary models}},
    year = {2020},
    journal = {Astrophysics A{\&}A},
    author = {Haldemann, Jonas and Alibert, Yann and Mordasini, Christoph and Benz, Willy},
    pages = {105},
    volume = {643},
    url = {https://doi.org/10.1051/0004-6361/202038367 https://www.aanda.org/articles/aa/pdf/2020/11/aa38367-20.pdf},
    doi = {10.1051/0004-6361/202038367}
}
\bibliographystyle{aasjournal}

\end{document}